# FlowMRI-Net: A Generalizable Self-Supervised 4D Flow MRI Reconstruction Network

Luuk Jacobs[1], Marco Piccirelli[2], Valery Vishnevskiy[1], Sebastian Kozerke[1]

[1] Institute for Biomedical Engineering, University and ETH Zurich, Zurich, Switzerland
[2] Department of Neuroradiology, University Hospital Zurich, Zurich, Switzerland

## Abstract

**Background:** Image reconstruction from highly undersampled 4D flow MRI data can be very time consuming and may result in significant underestimation of velocities depending on regularization, thereby limiting the applicability of the method. The objective of the present work was to develop a generalizable self-supervised deep learning-based framework for fast and accurate reconstruction of highly undersampled 4D flow MRI and to demonstrate the utility of the framework for aortic and cerebrovascular applications.

**Methods:** The proposed deep-learning-based framework, called FlowMRI-Net, employs physics-driven unrolled optimization using a complex-valued convolutional recurrent neural network and is trained in a self-supervised manner. The generalizability of the framework is evaluated using aortic and cerebrovascular 4D flow MRI acquisitions acquired on systems from two different vendors for various undersampling factors (R=8,16,24) and compared to state-of-the-art compressed sensing (CS-LLR) and deep learning-based (FlowVN) reconstructions. Evaluation includes an ablation study and a qualitative and quantitative analysis of image and velocity magnitudes.

**Results:** FlowMRI-Net outperforms CS-LLR and FlowVN for aortic 4D flow MRI reconstruction, resulting in significantly lower vectorial normalized root mean square error and mean directional errors for velocities in the thoracic aorta. Furthermore, the feasibility of FlowMRI-Net's generalizability is demonstrated for cerebrovascular 4D flow MRI reconstruction, where no FlowVN can be trained due to the lack of high-quality reference data. Reconstruction times ranged from 3 to 7 minutes on commodity CPU/GPU hardware.

**Conclusion:** FlowMRI-Net enables fast and accurate reconstruction of highly undersampled aortic and cerebrovascular 4D flow MRI, with possible applications to other vascular territories.

# Abbreviations

| | |
|---|---|
| 2D | two-dimensional |
| 3D | three-dimensional |
| 4D | four-dimensional |
| 5D | five-dimensional |
| MRI | magnetic resonance imaging |
| PI | parallel imaging |
| CS | compressed sensing |
| LLR | locally low rank |
| REF | reference image |
| CPU | central processing unit |
| GPU | graphical processing unit |
| DL | deep learning |
| CNN | convolutional neural network |
| RNN | recurrent neural network |
| VN | variational network |
| DC | data consistency |
| WA | weighted averaging |
| ECG | electrocardiogram |
| PPU | peripheral pulse unit |
| FOV | field of view |
| VENC | velocity encoding |
| TE | echo time |
| TR | repetition time |
| FA | flip angle |
| SNR | signal-to-noise ratio |
| CoW | circle of Willis |
| R | undersampling factor |
| FH | feet-head |
| RL | right-left |
| AP | anteroposterior |
| ROI | region of interest |
| MCA | middle carotid artery |
| ICA | internal carotid artery |
| PCA | posterior carotid artery |

# Introduction

4D flow magnetic resonance imaging (MRI) facilitates quantification of 4D (3D+time) blood flow dynamics [1], from which various hemodynamic parameters can be inferred such as wall shear stress [2], pressure gradients [3], and pulse wave velocity [4], with applications including the cardiovascular system (aorta, pulmonary arteries, abdomen, and liver), heart (atria, ventricles, and coronary arteries), and head/neck (carotid arteries and cerebral arteries and veins) [5]. However, clinical adaptation of 4D flow MRI [6], [7] has been hampered by its long scan time, particularly for smaller vasculatures that require higher spatial resolutions for accurate flow quantification.

Historically, scan times of MRI acquisitions have been successfully reduced by undersampling the acquired data and resolving the resulting aliasing artifacts using traditional reconstruction algorithms such as parallel imaging (PI) [8], [9], [10] and compressed sensing (CS) [11] methods, which exploit redundancies among multi-channel coils and compressibility of images, respectively. Although these methods could be used for frame-by-frame reconstruction of dynamic MRI, methods that also exploit temporal redundancies such as k-t BLAST (Broad-use Linear Acquisition Speed-up Technique) [12] and its temporally constrained extension k-t PCA (Principal Component Analysis) [13] have been shown to improve MRI reconstruction performance, including flow quantification [14], [15].

More recently, deep learning (DL)-based reconstructions have been demonstrated to enable even higher undersampling rates and/or faster reconstruction times, by exploiting redundancies that are implicitly learned from the acquired data [16]. Although a conceptually straightforward end-to-end mapping between undersampled k-space and fully-sampled image could be learned, this requires large amounts of paired data for training and does not take the known imaging physics into account [17]. Currently, the state-of-the-art is defined by unrolled optimization algorithms using physics-driven DL, pioneered by the variational network (VN) approach [18], which combines the expressiveness of DL with the robustness of traditional optimization algorithms [19]. Subsequently, FlowVN was developed for 4D flow MRI reconstruction, which trains an improved VN in a supervised manner using PI-reconstructed images as reference [20]. For aortic stenosis patients, it was shown that the relative error of velocity magnitudes for 10-fold retrospectively undersampled data was lower for FlowVN vs. CS-LLR (locally low rank), with reconstruction times of 21~s vs. 10~min, respectively. However, FlowVN's applications are limited to relatively small k-space matrices, such as its demonstrated aortic Cartesian 4D flow MRI use case, due to 1) high graphical processing unit (GPU) memory demands and 2) reliance on reference data.

Recent developments in DL training strategies can be used to overcome these limitations and improve generalizability. Firstly, memory-efficient learning strategies [21], such as gradient checkpointing [22], can be used to lower GPU memory demands that come with training unrolled networks and generalize applications to large-scale multidimensional data [19]. Secondly, self-supervised learning strategies can aid to generalize to applications where high-quality reference data is impractical or even infeasible to acquire, for example due to higher spatial and/or temporal resolution or multiple velocity encodings [23], [24], [25]. The recently proposed self-supervised learning via data undersampling (SSDU) [26] splits the undersampled data into two disjoint sets, one for network input and one for defining the training loss. SSDU and its multi-mask extension [27] have been successfully applied for reconstructions of various cardiac MRI applications [28], [29], [30], [31], but not for 4D flow MRI.

In the present work we propose FlowMRI-Net for generalizable 4D flow MRI reconstruction from highly undersampled data within clinically feasible time budgets. We demonstrate its generalizability obtained by modern memory-efficient and self-supervised learning strategies for aortic and cerebrovascular 4D flow MRI. Besides its improved generalizability, superior velocity quantification is achieved relative to state-of-the-art compressed sensing and variational network-based approaches via the novel exploitation of redundancies in velocity-encoding dimensions using complex-valued convolutions.

# Methods

The standard way to encode a time-resolved three-dimensional velocity-vector field is using a four-point referenced encoding scheme, with three scans encoding the velocity of each Cartesian direction independently in the phase of the signal using a bipolar gradient and one non-velocity-encoding scan measuring the background phase (number of velocity-encodings $N_V = 4$)[32]. The velocity field can then be extracted from the phase difference of these resulting complex-valued MR images $x \in \mathbb{C}^P$ ($P = N_V \times N_T \times N_X \times N_Y \times N_Z$), with number of cardiac bins $N_T$ and number of sampling points in $x, y$, and $z$ direction $N_X, N_Y$, and $N_Z$, respectively. Because of the high dimensionality of $x$, undersampling in k-space is required to make the acquisition time clinically feasible. Let $y_i \in \mathbb{C}^Q$ ($i \in \{1, \dots, N_C\}, Q \ll P$) represent the undersampled k-space measured by the $i$-th out of $N_C$ receiver coils. The forward MRI acquisition process can then be modelled as

$$y_i = E_i x = D\mathcal{F}S_i x, \tag{1}$$

with forward encoding operator $E_i$, consisting of the $i$-th coil sensitivity maps $S_i \in \mathbb{C}^{N_X \times N_Y \times N_Z}$, spatial Fourier transform $\mathcal{F}$, and undersampling mask given by the diagonal matrix $D$. To reconstruct $x$ from $y$, the resulting ill-posed inverse problem can be solved using the following unconstrained optimization:

$$\operatorname*{argmin}_{x} \frac{\lambda}{2} \sum_{i=1}^{N_C} \|D\mathcal{F}S_i x - y_i\|_2^2 + \mathcal{R}(x), \tag{2}$$

with weight $\lambda$ balancing the data consistency (DC) and regularization $\mathcal{R}$.

## MR image reconstruction using unrolled network

To solve Eq. 2 we use the variable splitting algorithm with quadratic penalties as proposed for VS-Net (variable splitting network) [33] by introducing two auxiliary variables $w_i$ and $z$ with quadratic penalty weights $\alpha$ and $\beta$, respectively:

$$\operatorname*{argmin}_{x,z,w_i} \frac{\lambda}{2} \sum_{i=1}^{N_C} \|D\mathcal{F}w_i - y_i\|_2^2 + \mathcal{R}(z) + \frac{\alpha}{2} \sum_{i=1}^{N_c} \|w_i - S_i x\|_2^2 + \frac{\beta}{2} \|x - z\|_2^2. \tag{3}$$

This formulation decouples the regularization from the DC term and prevents any dense matrix inversions later. To solve this multivariable optimization problem, alternate minimization over $x$, $w_i$, and $z$ is performed:

$$z^n = \operatorname*{argmin}_{z} \frac{\beta}{2} \|x^{n-1} - z\|_2^2 + \mathcal{R}(z), \tag{4a}$$

$$w_i^n = \operatorname*{argmin}_{w_i} \frac{\lambda}{2} \sum_{i=1}^{N_C} \|D\mathcal{F}w_i - y_i\|_2^2 + \frac{\alpha}{2} \sum_{i=1}^{N_c} \|w_i - S_i x^{n-1}\|_2^2, \tag{4b}$$

$$x^n = \operatorname*{argmin}_{x} \frac{\alpha}{2} \sum_{i=1}^{N_c} \|w_i^n - S_i x\|_2^2 + \frac{\beta}{2} \|x - z^n\|_2^2, \tag{4c}$$

with $n \in \{1, \dots, N\}$ denoting the $n$-th iteration and $x^0$ being initialized as a zero-filled reconstruction, where Eq. 4a is the proximal operator of the prior $\mathcal{R}$ and Eqs. 4b and 4c have closed-form solutions:

$$z^n = prox_{\mathcal{R}}(x^{n-1}), \tag{5a}$$

$$w_i^n = \mathcal{F}^{-1}\left(\left(\lambda D^T D + \alpha I\right)^{-1}\left(\lambda D^T y_i + \alpha \mathcal{F}S_i x^{n-1}\right)\right), \tag{5b}$$

$$x^n = \left(\alpha \sum_{i=1}^{N_c} S_i^H S_i + \beta I\right)^{-1} \left(\alpha \sum_{i=1}^{N_c} S_i^H w_i^k + \beta z^n\right). \tag{5c}$$

Here, Eq. 5a can be defined as a *learnable* denoising operation using a neural network (*denoising block*). Note that the inversions in Eqs. 5b and 5c are performed on diagonal matrices, making them element-wise operations. When defining noise level $\nu^n = \frac{\lambda}{\lambda + \alpha}$ per iteration $n$, Eq. 5b turns into a coil-wise weighted averaging that sums the measured k-space $\nu^n y_i$ and reconstructed k-space $(1 - \nu^n)\mathcal{F}S_i x^{n-1}$ for sampled ($D_{jj} = 1$) points and copies the reconstructed k-space $\mathcal{F}S_i x^{n-1}$ for unsampled ($D_{jj} = 0$) points (*DC block*). Similarly, when defining weighting $\mu^n = \frac{\alpha}{\alpha + \beta}$ per iteration $n$ and assuming normalized $S_i$, Eq. 5c turns into a weighted averaging that sums the output of the denoising block $\mu^n \sum_{i=1}^{N_c} S_i^H w_i^n$ and

DC block $(1 - \mu^n)\boldsymbol{z^n}$ (*weighted-averaging (WA) block*). The denoising, DC, and WA blocks are repeated for a fixed number of $N$ units in the unrolled network, where both $v^n$ and $\mu^n$ are real-valued learnable parameters per unit $n$ and are passed through a Sigmoid function $\sigma$ to guarantee proper averaging.

### Data Acquisition

**Aortic data** [34] were acquired in 15 volunteers (age=28.1±3.9 years, m/f=2/3) after written informed consent and according to institutional and ethical guidelines on a Philips Ambition 1.5T system (Philips Healthcare, Best, the Netherlands) with a 28-channel cardiac coil using a retrospectively electrocardiogram (ECG)-triggered four-point referenced phase-contrast gradient-echo sequence during free breathing. The sagittal oblique field-of-view (FOV) = 360~mm x 228-298~mm x 60~mm (minimized in phase-encoding direction to maximize scan efficiency whilst preventing fold-over) covered the thoracic aorta, using an acquired spatial resolution of 2.5~mm x 2.5~mm x 2.5~mm, an acquired temporal resolution of 48.9~ms, velocity encoding (VENC) = 150~cm/s, echo time (TE) = 3.1~ms, repetition time (TR) = 5.4~ms, and flip angle (FA) = 15~°. 3D Cartesian elliptical pseudo-spiral Golden angle data were acquired for approximately one hour per volunteer, from which three equally populated respiratory bins were extracted using the Philips VitalEye camera [35], where only the end-expiratory bin was considered in this work as this is the most reproducible state during free breathing, resulting in an effective undersampling factor between 3 and 4, depending on heartrate and FOV. Binning into three respiratory states was empirically found to be the optimal trade-off between limiting respiratory motion and the effective undersampling factor. A CS-based reconstruction (please refer to Experiment and evaluation for more details) of these relatively densely sampled data were used for evaluation. Due to the Golden angle sampling, paired prospectively-undersampled scans could be extracted from the acquisition by omitting the last number of readouts, depending on the desired undersampling factor. As an additional reference, a fully sampled two-point 2D referenced phase-contrast gradient-echo was acquired for each of the Cartesian directions during breath-holding. The axial slice with a FOV = 350~mm x 200-425~mm (maximized in phase-encoding direction to maximize the signal-to-noise ratio (SNR)) cut the ascending and descending aorta at the level of the right pulmonary trunk, using an acquired spatial resolution of 2.5~mm x 2.5~mm x 8~mm, an a temporal resolution of 45.8~ms, VENC = 150~cm/s, TE = 2.6~ms, TR = 4.5~ms, and FA = 15~°.

**Cerebrovascular data** [34] were acquired in 10 volunteers (age = 25.2 ± 3.1 years, m/f = 3/2) after written informed consent and according to institutional and ethical guidelines on a Siemens Vida 3T system (Siemens Healthineers, Erlangen, Germany) with a 64-channel head-neck coil using a retrospectively peripheral pulse unit (PPU)-triggered four-point referenced phase-contrast gradient-echo sequence. The axially oblique FOV = 240~mm x 183~mm x 64~mm covered the circle of Willis (CoW) and the confluence of sinuses, using an acquired spatial resolution of 0.8~mm x 0.8~mm x 0.8~mm, an acquired temporal resolution of 55.8~ms, VENC = 100~cm/s, TE = 4.14~ms, TR = 6.98~ms, and FA = 13~°. Each volunteer was scanned using a 3D Cartesian 8-fold pseudo-random sampling pattern, taking approximately 19 minutes depending on heartrate. One volunteer was also scanned using a 2-fold generalized autocalibrating partial parallel acquisition (GRAPPA) [9] with 54 reference lines and a reduced FOV = 240~mm x 171~mm x 48~mm, requiring an hour, which was used as a reference to test reconstruction fidelity.

### Data Pre-Processing

The raw aortic and cerebrovascular data were parsed using PRecon (GyroTools LLC, Zurich, Switzerland) and twixtools [36], respectively. The complex-valued Cartesian k-space data $\boldsymbol{y}$ were acquired with a fully sampled readout dimension, allowing the k-space to be split into $N_X$ subvolumes. Estimation of coil sensitivities $\boldsymbol{S}$ from the time-averaged k-space data and subsequent compression into $N_C = 10$ virtual coils were performed using the Berkeley Advanced Reconstruction Toolbox (BART) [37]. All velocity-encodings were acquired in an interleaved manner and jointly reconstructed by using them as feature channels of the network input, which allows their redundancies to be exploited [38]. Note that the real and imaginary components do not need to be separated in the feature channel dimension because

FlowMRI-Net is complex-valued, which was found to be particularly beneficial for phase-focused applications [39]. Because only a small subset of each cerebrovascular volume contains voxels with motion/flow, an automatic sliding threshold segmentation [40] was performed on the time-averaged complex difference volumes [41] and only slices with more than 0.2% vessel content were considered for training.

### Network Architecture And Training

The proposed unrolled network architecture and its self-supervised training scheme can be seen in Fig. 1a (see "MR image reconstruction using unrolled networks" section). For the denoising block we adopted a variation of the convolutional recurrent neural network (CRNN)-MRI [42], with four bidirectional CRNN (BCRNN) layers that evolve recurrence over both temporal and iteration dimensions and one CNN layer. We used 2D complex-valued convolutions applied over spatial $y, z$ dimensions with kernel size $k = 3$, number of filters $f = 25$, and $N = 10$ unrolling units that share their weights and biases. The complex-valued phase-preserving modReLU, a modified rectified linear unit (ReLU) [43] proposed for complex-valued RNNs, is used as activation function [44]:

$$modReLU(z) = \begin{cases} (|z| + b)\frac{z}{|z|}, & if \ |z| + b \geq 0 \\ 0, & if \ |z| + b < 0 \end{cases}, \quad (6)$$

where $b$ is a real-valued bias parameter learnable for each feature map. The input of this denoising block is summed with its output by a residual connection. In parallel, the input is fed to the DC block, whose output is merged with the output of the denoising block in the WA block. The learnable noise levels $\nu^n$ (for DC layers) and learnable weights $\mu^n$ (for WA layers) for each unrolling unit n can be seen in Fig. 1b. Similar patterns can be seen for aortic and cerebrovascular reconstructions at different undersampling factors: an initial decrease in $\nu^n$ with every unrolling unit resulting in an increased weight of the acquired samples during DC, with more weight on the predicted samples for the final layers, and a gradual decrease of $\mu^n$ in the WA layers resulting in increasingly more weight on the DC output compared to the denoising block output.

To lower GPU memory demands at the cost of training time, gradient checkpointing was used per unrolling unit to recompute intermediate activations during backpropagation rather than saving them all during the forward pass [22]. Furthermore, tensors that were saved in the forward pass can be offloaded and stored on the central processing units (CPUs) [45].

Given the acquired k-space $\boldsymbol{y_\Omega}$ and estimated coil sensitivities $\boldsymbol{S}$, the sampled locations $\Omega$ are partitioned into two disjoint sets $\Theta$ and $\Lambda$. The zero-filled image of the first set, computed using the adjoint $\boldsymbol{E_\Theta^H}$ of the forward operator $\boldsymbol{E_\Theta}$, was used as network input and its k-space $\boldsymbol{y_\Theta}$ was used for DC. The k-space of the second set $\boldsymbol{y_\Lambda}$ was used for defining the loss. The partitioning was performed using uniform random sampling with an 80/20 split, where set $\Theta$ always contained the circular center of k-space with radius 3 for training stability. Additionally, samples that differ only in velocity-encoding compared to samples in $\Lambda$, were also included into $\Lambda$ as this was empirically found to reduce reconstruction artefacts. Note that during inference, no partitioning is necessary and $\boldsymbol{y_\Omega}$ can be used as network input. In the original multi-mask SSDU setup [27], each partition was randomly repeated for a fixed number of times per measurement, where too many partitions (considered to be data augmentations) resulted in overfitting [46]. However, such overfitting was never observed in any of our experiments, attributable to the relatively small number of learnable parameters ($\sim 10^4$) compared to the number of data samples per measurement ($\sim 10^6$). Accordingly, the partitioning was randomized for each iteration. For the training loss, reconstructed image $x_\Theta^N$ was converted to k-space $\boldsymbol{y_\Theta^N} = \boldsymbol{E_\Lambda} x_\Theta^N = \boldsymbol{D_\Lambda \mathcal{F} S} x_\Theta^N$ and compared to the left-out set $\boldsymbol{y_\Lambda}$ using a point-wise normalized L1-L2 loss [26]:

$$\mathcal{L}(\boldsymbol{y_\Lambda}, \boldsymbol{y_\Theta^N}) = \frac{\|\boldsymbol{y_\Lambda} - \boldsymbol{y_\Theta^N}\|_2}{\|\boldsymbol{y_\Lambda}\|_2} + \frac{\|\boldsymbol{y_\Lambda} - \boldsymbol{y_\Theta^N}\|_1}{\|\boldsymbol{y_\Lambda}\|_1}. \quad (7)$$

This self-supervised loss was optimized using the Adam algorithm [47] with a batch size of 1 and a learning rate of $5 \times 10^{-4}$ with a cosine annealing to 0 over 50 epochs (equivalent to $5.9 \times 10^4$ and $6.5 \times 10^4$ iterations for aortic and cerebrovascular data, respectively). All training was performed using

Pytorch 2.3.1 [48] with Pytorch Lightning 2.2.0 [49] on an Nvidia Titan RTX GPU with 24GB memory and a 32-core Intel Xeon Gold 6130 CPU running at 2.10 GHz with 188GB memory. For the aortic data, the 15 volunteers were split into 9, 1, and 5 for training, validation, and testing, respectively. Hyperparameters were optimized using the validation set of the aortic data and were kept the same for reconstructions of the cerebrovascular data, where only 1 of the 10 volunteers had a reference acquisition (see Data acquisition). Consequently, its 10 volunteers were split into 9 for training and 1 for testing. FlowVN and FlowMRI-Net were trained separately for aortic and cerebrovascular reconstructions and for each undersampling factor.

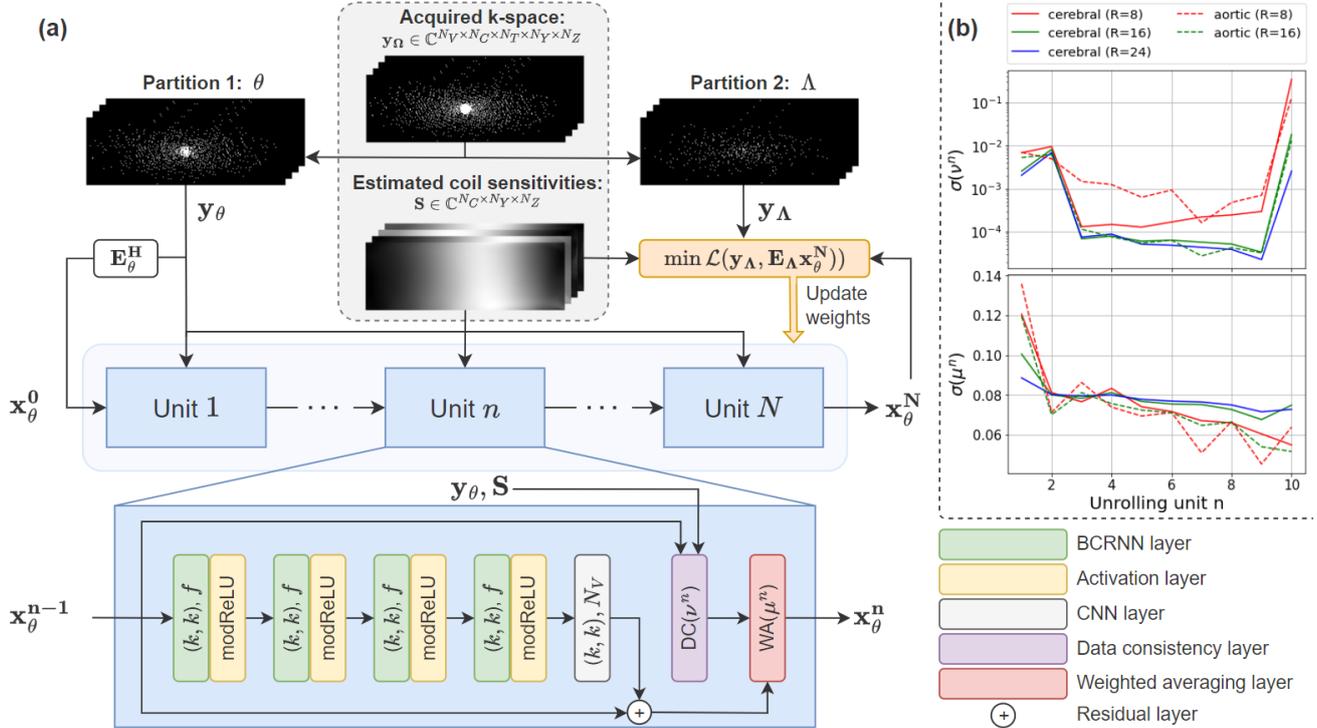

**Figure 1: An overview of the self-supervised learning scheme of FlowMRI-Net.** (a) The acquired k-space data $y_\Omega$ are randomly partitioned into disjoint sets $\Theta$ and $\Lambda$. Partition $\Theta$ is used as input for the neural network with $N$ unrolling units, exploiting spatiotemporal redundancies in bidirectional recurrent neural network (BCRNN) and convolutional neural network (CNN) layers (with kernel size $k$ and number of filters $f$ or $N_V$) and enforcing point-wise data consistency in data consistency (DC) layers, which are combined in the weighted averaging (WA) layer. The resulting reconstruction $x_\theta^N$ is transformed to k-space and compared to the heldout second partition $y_\Lambda$ using a normalized L1-L2 loss. (b) The Sigmoid $\sigma(.)$ of the learnable noise level $\nu^n$ (for DC layers) and learnable weight $\mu^n$ (for WA layers) are shown for each of the N=10 unrolling units trained using aortic and cerebrovascular data for different undersampling factors (R).

### Experiments And Evaluation

The aortic data were reconstructed for two prospective undersampling factors ($R = 8$ and 16, with acquisition times of 19 and 7 minutes assuming a heart rate of 60 beats per minute, respectively) using FlowMRI-Net, FlowVN, and a GPU-accelerated CS using locally low rank (LLR) [50] regularization (CS-LLR). An ablation study was performed to investigate the importance of FlowMRI-Net's joint velocity-encoding reconstruction and complex-valued convolutions by including two additional versions of FlowMRI-Net: 1) single-encoding reconstruction (randomized each iteration as for FlowVN) and 2) using real-valued convolutions (real and imaginary components added to input feature dimension as for FlowVN). The hyperparameters of FlowVN and CS-LLR were optimized using the validation set. FlowVN was trained using the CS-LLR reconstruction of the relatively densely sampled ($R = 4$) acquisition for the loss calculations of its supervised training. Although the original FlowVN implementation used retrospectively-undersampled data from its reference scans with additional k-space noise as network input during training, we directly input the prospectively-undersampled scans, resulting in less overfitting and a fairer comparison to CS-LLR and FlowMRI-Net. For testing, all aortic reconstructions were compared relative to CS-LLR ($R = 4$). Quantitative metrics were computed for the systolic phases (here defined as the third to fifth cardiac bin) separately and included the normalized peak-systolic root-mean-square error (nRMSE) of the image magnitude:

$$\text{nRMSE}_m(a, a^*) = \sqrt{\frac{\sum |a - a^*|^2}{\sum |a^*|^2}}, \tag{8}$$

for magnitude images $a$ and $a^*$, the vectorial nRMSE of velocities inside a region-of-interest (ROI) in the thoracic aorta:

$$\text{nRMSE}_v(\boldsymbol{u}, \boldsymbol{v}) = \sqrt{\frac{\sum_{i \in ROI} \|\boldsymbol{u}_i - \boldsymbol{v}_i\|^2}{\sum_{i \in ROI} \|\boldsymbol{v}_i\|^2}}, \tag{9}$$

for vector fields $\boldsymbol{u}$ and $\boldsymbol{v}$, and the mean directional error (mDirErr) [51] of velocities in the ROI:

$$\text{mDirErr}(\boldsymbol{u}, \boldsymbol{v}) = \frac{1}{|ROI|} \sum_{i \in ROI} 1 - \frac{|\boldsymbol{u}_i \cdot \boldsymbol{v}_i|}{\|\boldsymbol{u}_i\| \|\boldsymbol{v}_i\|}. \tag{10}$$

The $\text{nRMSE}_m$, $\text{nRMSE}_v$, and mDirErr values of the five reconstruction methods (CS-LLR, FlowVN, and 3 versions of FlowMRI-Net) were visualized using boxplots and their statistical differences were compared using non-parametric two-sided Mann-Whitney U tests with a significance level of $p \leq 0.05$.

Furthermore, the velocity curves during peak systole and peak velocity magnitude in the ascending aorta of the reconstructions were compared to the fully sampled 2D breath-hold reference (2D REF) acquisitions with velocity encoding in feet-head (FH), right-left (RL), and anteroposterior (AP) direction. The peak velocity magnitude was computed by determining the spatiotemporal coordinate with the highest velocity in FH direction within an eroded ROI and extracting the velocities in RL and AP direction at that same coordinate. Velocity curves were determined by tracing that coordinate over time for each encoding direction. The time-dependent 2D cross-sections of the ascending aorta were semi-automatically segmented using ITK-SNAP's snake evolution algorithm [52] and the time-dependent 3D intra-vessel volumes of the thoracic aorta were automatically segmented using an in-house 3D nnU-net [53], starting at the sinotubular junction and excluding the brachiocephalic, left common carotid, and left subclavian arteries [54].

The 8-fold undersampled cerebrovascular data were further reduced using random Gaussian sampling ($R = 16$ and 24, with acquisition times of 9.5 and 4.75 minutes assuming a heart rate of 60 beats per minute, respectively) and were reconstructed using FlowMRI-Net and CS-LLR. Because high-quality reference data are unavailable, no FlowVN could be trained and reconstructions were predominantly qualitatively compared to the 2-fold undersampling GRAPPA reconstruction for the testing set, where the SNR was quantitatively compared for patches placed in homogeneous parts of the brainstem and white matter of the image magnitude ($a_{patch}$), assuming Gaussian noise:

$$SNR(a_{patch}) = 20 \log_{10} \frac{mean(a_{patch})}{std(a_{patch})}, \tag{11}$$

where $std(.)$ computes the standard deviation. The GRAPPA reconstruction was rigidly registered to the FlowMRI-Net reconstruction using SimpleITK's 3D Euler transform [55]. The 3D segmentations for the velocity curves of the left and right middle cerebral arteries (LMCA/RMCA) at the M1 level, left and right internal carotid arteries (LICA/RICA) at C3-C4 level, and left and right posterior cerebral arteries (LPCA/RPCA) at the P2b level were semi-automatically segmented using ITK-SNAP's snake evolution algorithm [52].

Upon reconstructions, magnitude inhomogeneities were corrected using N4ITK bias field correction [56], concomitant fields for the aortic data were corrected according to [57], eddy currents were corrected using a third-order polynomial fitting to a stationary tissue volume or slice using M-estimate SAmple Consensus (MSAC) [58], and velocity aliasings were corrected using 4D Laplacian phase unwrapping [59], [60].

# Results

Since both FlowVN and FlowMRI-Net are unrolled networks with shared weights between unrolling iterations, the total number of learnable parameters is relatively low, which is why little data is needed for training (Table 1). Although training times are high for these DL-based reconstructions, training had to be performed only once per undersampling factor. Inference times for FlowMRI-Net ranged between 1 (aorta) and 7 min (cerebrovascular) on CPU/GPU commodity hardware. The increased training and inference times of FlowMRI-Net compared to FlowVN can be attributed to the use of all cardiac bins and the relatively time-consuming forward and backward passes of the complex-valued BCRNN layers.

| Anatomy | Method | Number of parameters | Training time | Inference time |
|---|---|---|---|---|
| Aorta | CS-LLR | 2 | - | 1~min |
| | FlowVN | 63460 | 27.5~h | 1~min |
| | FlowMRI-Net | 64099 | 100.6~h | 3~min |
| Cerebrovascular | CS-LLR | 2 | - | 8~min* |
| | FlowMRI-Net | 64099 | 129.2~h | 7~min |

**Table 1: Model complexities, training times, and typical inference times for reconstruction of a $N_V \times N_T \times N_X \times N_Y \times N_Z$ dataset.** All trainings and inferences were performed on a Nvidia Titan RTX GPU and a 32-core Intel Xeon Gold 6130 CPU. Note that the exact inference times depend on the number of cardiac bins ($N_T$), which in turn depend on the heart rate during acquisition, and, for aortic data, also on the variable number of phase-encodes in AP direction ($N_Y$). *The four velocity-encodings had to be reconstructed separately due to GPU memory limitations

The quantitative metrics for aortic reconstructions are summarized in Fig. 2, with FlowVN resulting in significantly lower $nRMSE_m$ values compared to CS-LLR and FlowMRI-Net, while $nRMSE_v$ and $mDirErr$ values were similar for all reconstructions at R=8. FlowMRI-Net (with joint reconstruction of segments and complex-valued convolutions) resulted in the lowest $nRMSE_v$ and $mDirErr$ values for reconstructions at R=16. Image magnitudes, velocity magnitudes, and their errors of an exemplary healthy volunteer are shown at peak systole in Figs. 3 and 4 for 8-fold and 16-fold prospective undersampling, respectively.

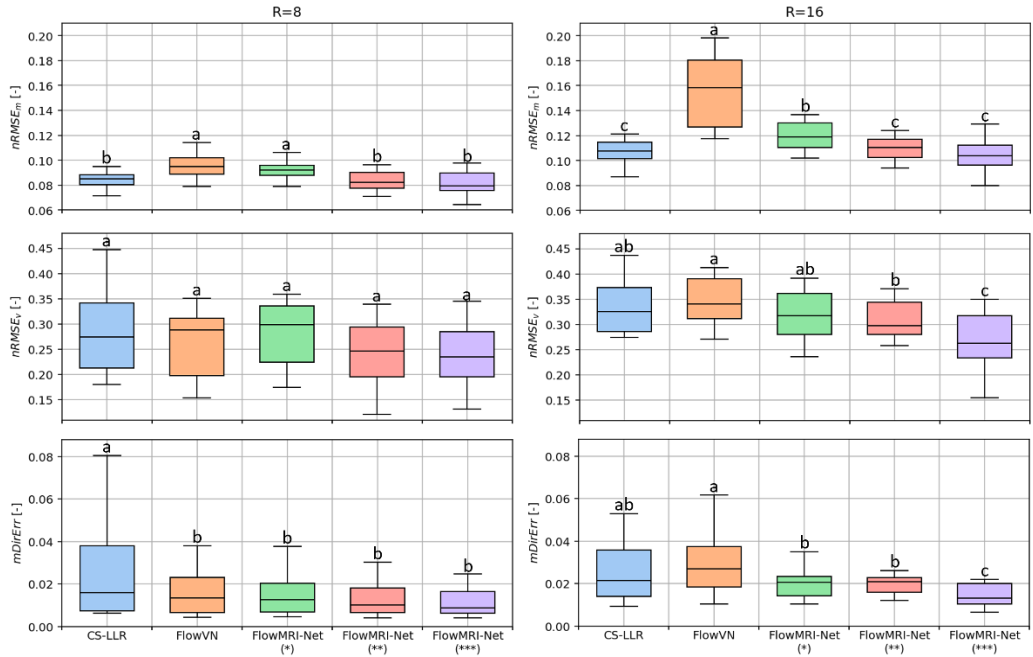

**Figure 2: Comparison of quantitative metrics for aortic reconstructions.** Metrics $nRMSE_m$, $nRMSE_v$, and $mDirErr$ are compared for prospective undersampling factors R=8,16 and statistically compared. Statistical significance is denoted using letters, where methods sharing a letter are not statistically significantly different. *Separate reconstruction of velocity-encodings. **Real-valued convolutions. ***Proposed: joint reconstruction of velocity-encodings with complex-valued convolutions.

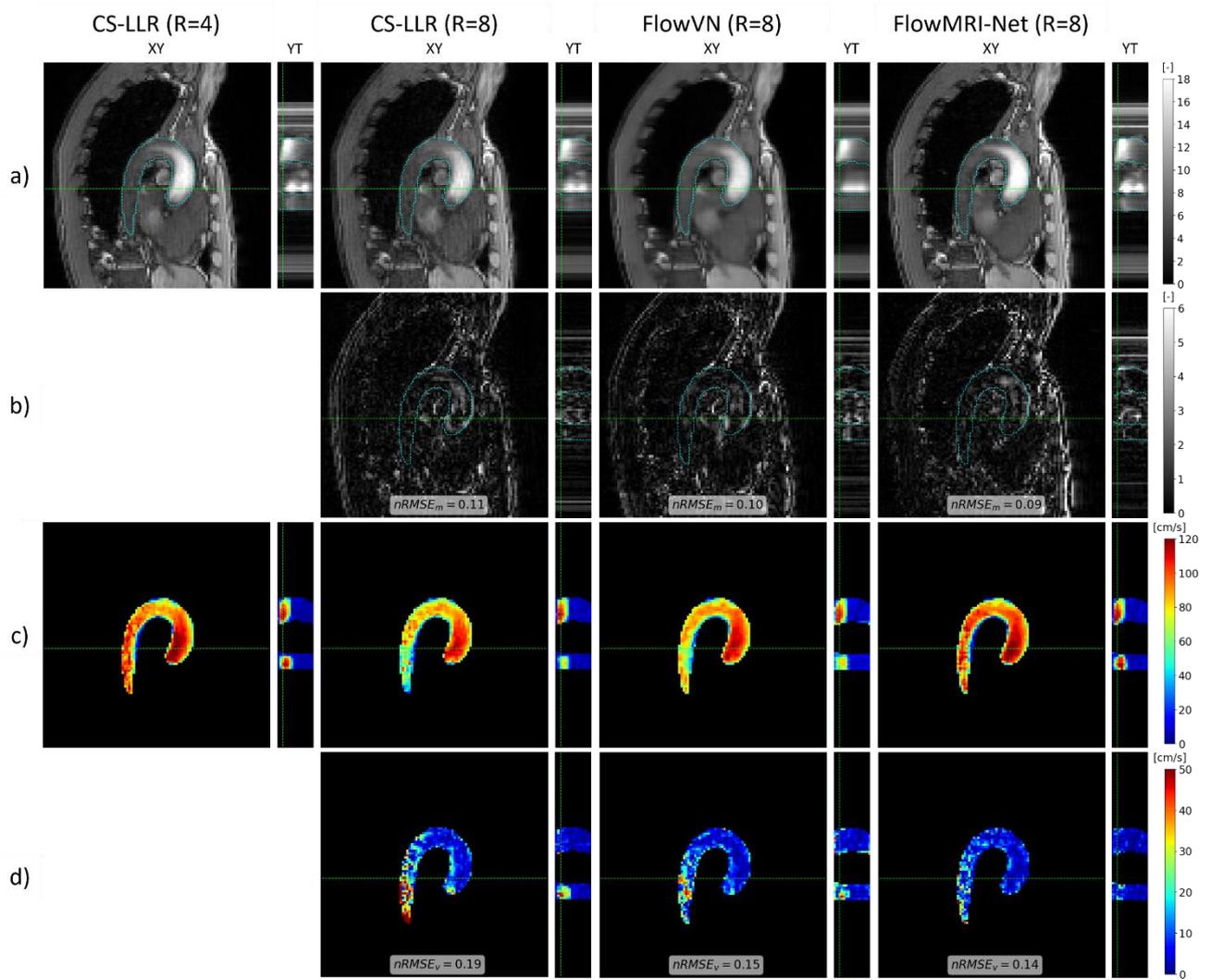

**Figure 3: Comparison of reconstruction methods for 8-fold prospectively undersampled aortic data at peak systole.** Image magnitudes (a), absolute image magnitude differences with CS-LLR (R=4) reconstruction (b), velocity magnitudes (c), and velocity magnitude differences with CS-LLR (R=4) reconstruction (d) are shown for a sagittal slice (XY) at peak systole and an AP column over time (YT). Corresponding slice locations and delineation of the aorta are illustrated with green and blue dashed lines, respectively.

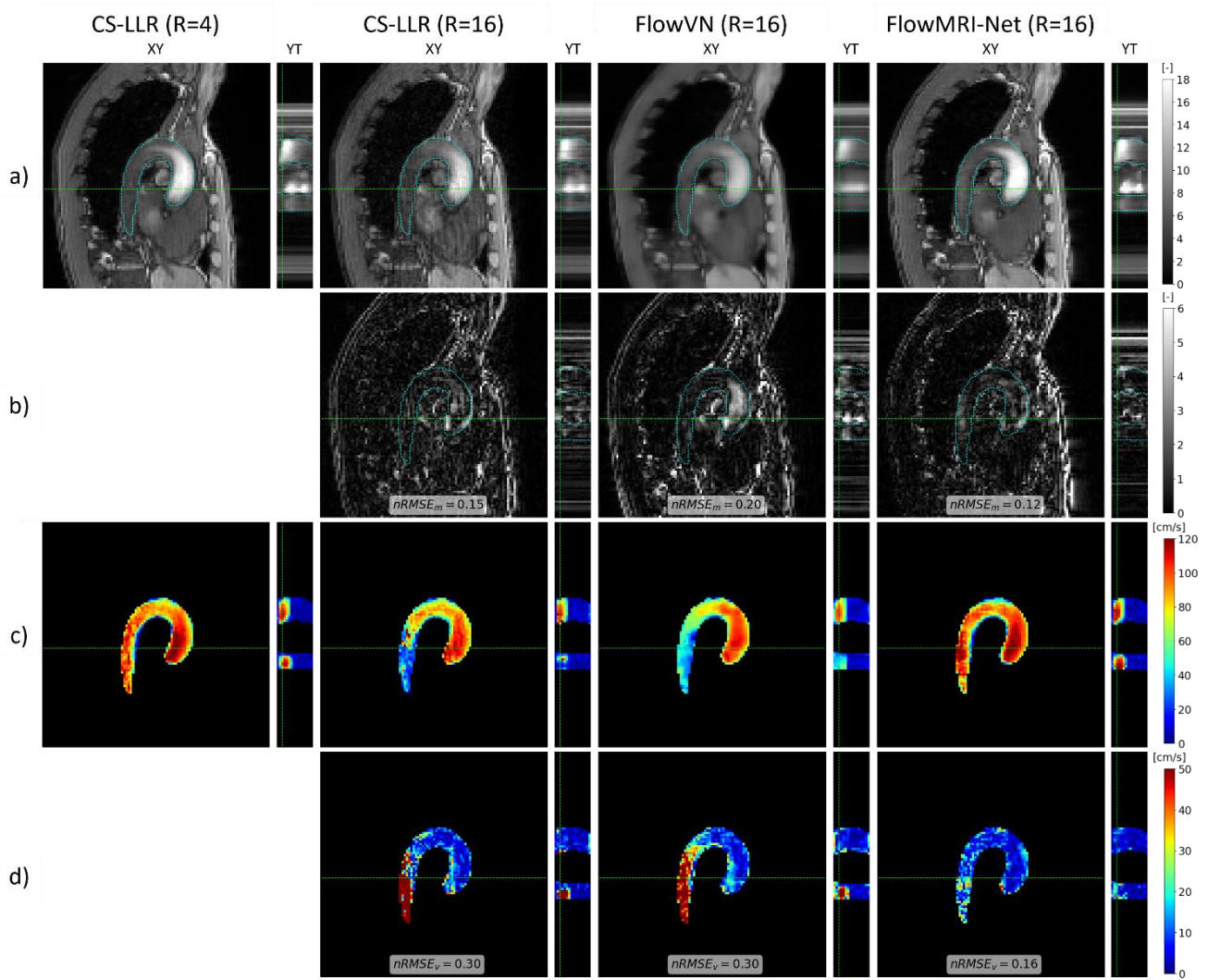

**Figure 4: Comparison of reconstruction methods for a 16-fold prospectively undersampled aortic data at peak systole.** Image magnitudes (a), absolute image magnitude differences with CS-LLR (R=4) reconstruction (b), velocity magnitudes (c), and velocity magnitude differences with CS-LLR (R=4) reconstruction (d) are shown for a sagittal slice (XY) at peak systole and an AP column over time (YT). Corresponding slice locations and delineation of the aorta are illustrated with green and blue dashed lines, respectively.

A comparison of peak velocities in the ascending aorta at the level of the right pulmonary artery between the fully sampled 2D REF and CS-LLR, FlowVN, and FlowMRI-Net reconstructions can be seen in Fig. 5 for undersampling factors R=8 (upper row) and R=16 (bottom row). In FH direction, an underestimation is observed for CS-LLR and FlowVN reconstructions at both undersampling factors while FlowMRI-Net accurately captures peak flow. In RL and AP directions, all three reconstruction methods showcase the correct velocity profile, but are relatively noisy.

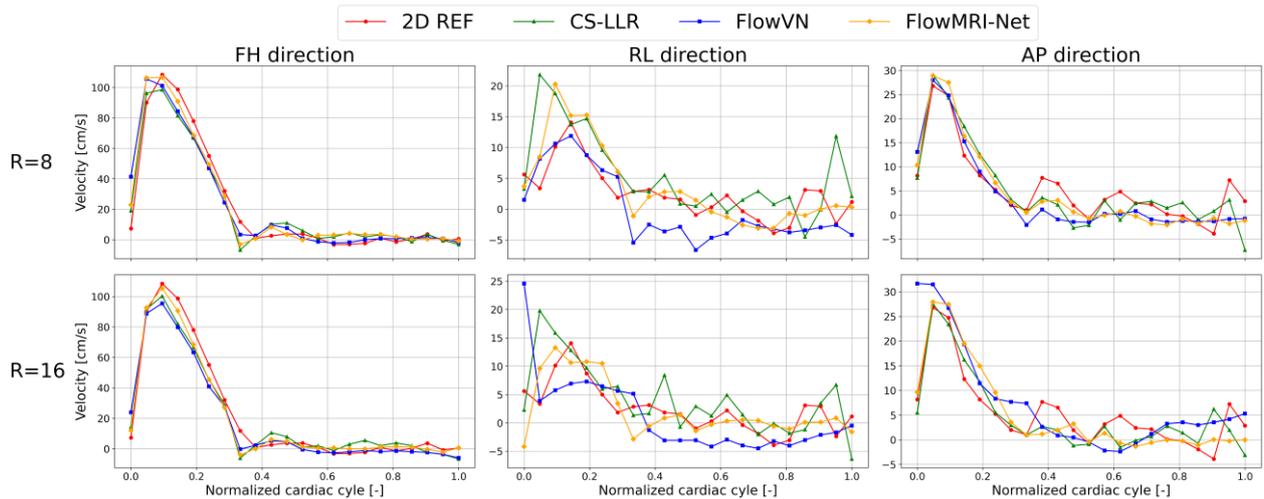

**Figure 5: Comparison of peak velocity curves in the ascending aorta.** CS-LLR, FlowVN, and FlowMRI-Net reconstructions of prospectively undersampled data, and 2D reference (REF) from fully sampled data acquired during breath-hold, for feet-head (FH), right-left (RL), and anterior-posterior (AP) directions at different undersampling factors (R).

Reconstructions of a cerebrovascular acquisition at undersampling factors R=8, 16, and 24 can be seen in Fig. 6. Visually, the image magnitude of the frame-by-frame reconstructed GRAPPA reference is noisier than the CS-LLR and FlowMRI-Net reconstruction at higher undersampling factors, which is reflected by the lower SNR, particularly in the brain stem furthest away from the coils. Undersampling artifacts in the form of noise-like structures are prominent in the center of the image magnitude for the CS-LLR reconstruction at R=24, with visible underestimation of velocity magnitudes at all undersampling factors, particularly in the left and right internal carotid arteries (ICA) and middle cerebral arteries (MCA). In contrast, the proposed FlowMRI-Net demonstrates better agreement of velocity magnitudes with the reference GRAPPA reconstruction, even up to the highest undersampling factor (R=24). These observations extend to the peak velocity curves for the left and right MCA and ICA, as seen in Fig. 7, where FlowMRI-Net shows superior agreement with the GRAPPA reconstruction compared to CS-LLR, which tends to underestimate peak velocity magnitudes. CS-LLR underestimates and FlowMRI-Net overestimates the peak velocity in the left and right PCA at all undersampling factors.

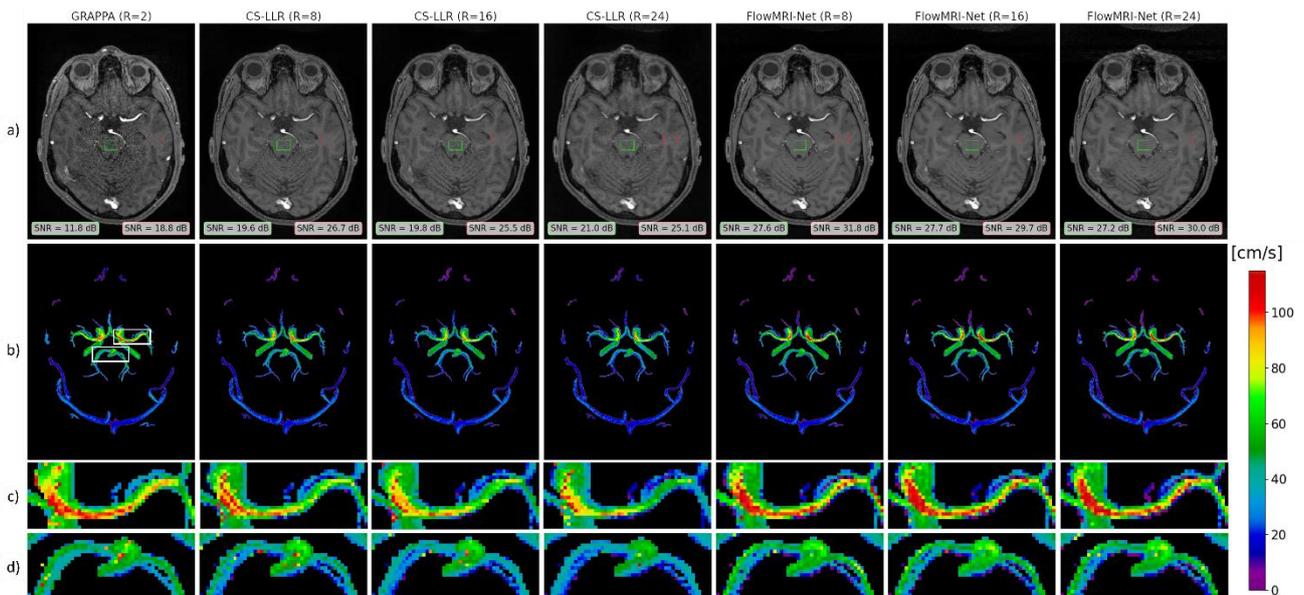

**Figure 6: Comparison of reconstruction methods for cerebrovascular data for various undersampling factors (R).** Image magnitudes (a), where the signal-to-noise ratio (SNR) has been computed for patches in the brain stem (green) and white matter (red), and maximum intensity projections of velocity magnitudes (b), with zoomed-in sections of the white rectangles on the left middle carotid artery (LMCA) (c) and left and right posterior carotid artery (LPCA/RPCA) (d).

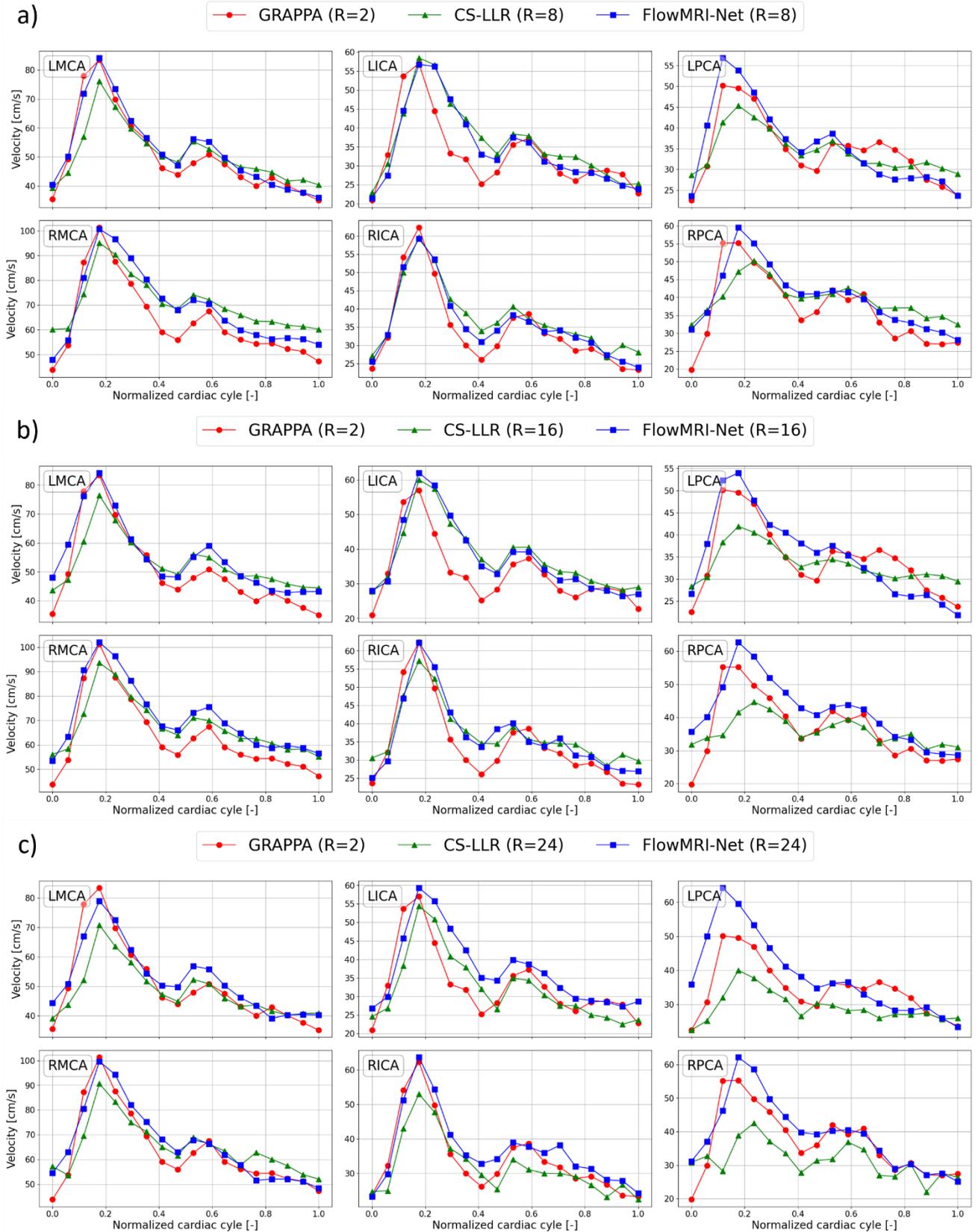

**Figure 7: Comparison of peak velocity curves for different vessels in cerebrovascular reconstructions at different undersampling factors (R).** The spatial coordinate with peak velocity magnitude is determined based on R=8 (a) and is traced for R=16 (b) and R=24 (c), all showing the left and right middle carotid artery (MCA), internal carotid artery (ICA), and posterior carotid artery (PCA).

# Discussion

In this work FlowMRI-Net has been proposed, showing improved reconstruction quality over FlowVN and CS-LLR for aortic 4D flow MRI and demonstrating generalizability for cerebrovascular 4D flow MRI, all with clinically feasible reconstruction times. With FlowMRI-Net several improvements compared to FlowVN are introduced: 1) a self-supervised learning strategy that does not rely on the availability or quality of an acquired reference dataset [27], 2) a memory-efficient learning strategy that lowers GPU memory demands 10-fold [22], 3) exploitation of the redundancies between velocity encodings, 4) preservation of the complex representation of the MRI data using a complex-valued network, and 5) recurrent convolutions that effectively propagate information across unrolling iterations and time, resulting in more complex feature representations for the same number of learnable parameters [42].

The superior reconstruction quality of prospectively undersampled aortic 4D flow MRI for R = 8 and 16 was demonstrated both quantitatively (Fig. 2) and qualitatively (Figs. 3 and 4), with FlowMRI-Net recovering more accurate structural details and hemodynamic features in the image magnitudes and velocity magnitudes compared to CS-LLR and FlowVN reconstructions. Particularly the superior recovery of temporal dynamics by FlowMRI-Net compared to CS-LLR and FlowVN can be appreciated in the YT planes of the image and velocity magnitudes. FlowMRI-Net's complex-valued convolution and exploitation of redundancies between velocity encodings result in improved $nRMSE_v$ and $mDirErr$ values for aortic reconstructions at R=16. Moreover, a clear underestimation of velocity magnitudes can be seen during peak systole by CS-LLR and FlowVN, particularly in the descending aorta where SNR is lower due to the higher distance to the coils.

The considerable spatiotemporal blurring that is seen for FlowVN reconstructions in this work was not reported in the original paper [20]. We hypothesize that this difference comes from the change in training data, specifically by switching from retrospective to prospective undersampling. Training on retrospectively undersampled data guarantees the input k-space data to be consistent (for example in terms of motion or gradient-induced field drifts) with the reference k-space data when computing the supervised loss. For prospectively undersampled data this consistency cannot be ensured, with a resulting mismatch, possibly leading to spatiotemporal blurring.

Compared to the fully sampled 2D breath-hold scans, FlowMRI-Net accurately captures peak velocity in FH direction, as opposed to CS-LLR and FlowVN, which both underestimate peak velocity (Fig. 5). Velocities in RL and AP direction remain noisy, which can be explained by the relatively high VENC = 150~cm/s compared to the maximum velocities that occur in those two directions for non-pathological flow ($v_{max}$ < 50~cm/s).

For cerebrovascular reconstructions, using a two-fold GRAPPA reconstruction as reference, we showed that FlowMRI-Net resulted in improved noise reduction in image magnitudes compared to CS-LLR (Fig. 6), whilst accurately capturing peak velocity magnitudes in the left and right MCA and ICA, even up to an undersampling factor of R=24 (Fig. 7). The velocities in the left and right PCA remain challenging to recover for both CS-LLR and FlowMRI-Net even at R=8 due to the relatively small vessel diameter (2.7 +/- 0.04~mm for the P2 segment that we considered) [61] compared to the spatial acquisition resolution of 0.8~mm x 0.8~mm x 0.8~mm, making flow quantification extremely sensitive to noise and partial volume effects, which may explain the relatively large under- and overestimation of peak velocities.

## Limitations

For the one-hour aortic reference scan, possible bulk motion and inevitably imperfect periodicity of cardiac and respiratory motion during acquisition can be expected to result in artefacts, limiting the strength of the validation but in-turn also supporting the use of self-supervised learning. In the present study, breathing motion during aortic acquisitions was tracked using the Philips VitalEye camera [35] and only the end-expiration state was considered, which reduces imaging efficiency and prohibits measurement of respiration-resolved flow dynamics (5D flow MRI) [25], [62]. Although the respiratory bins can be reconstructed separately, future work should investigate motion-informed reconstruction to exploit redundancies between breathing states, using learning-based [63] or conventional registration-based [64], [65] motion estimation. Additionally, FlowMRI-Net was trained per undersampling factor in the present study, and improved generalizability can be achieved by training a single FlowMRI-Net with DC and WA block conditioned on the undersampling factor as with FlowVN [20]. Furthermore, considering organ motion, time-resolved sensitivity maps may improve reconstruction quality compared to the current static maps. Finally, the generalizability to pathological flow dynamics, such as for aortic stenoses, should be investigated.

For the cerebrovascular acquisitions, the CS-like k-space sampling implemented on the Siemens system was not optimized and may be suboptimal for FlowMRI-Net. In brief, the sampling pattern has non-random patterns that cause higher lateral peaks in the corresponding point-spread function (i.e. higher coherence) compared to truly pseudo-random sampling, resulting in structured undersampling artifacts that are undesirable in our context [11]. Moreover, the corners of k-space, which are less critical, were sampled relatively densely; a proper elliptical shutter can increase imaging efficiency. Furthermore, it is currently unclear whether incoherent/complementary sampling between velocity-encodings is desirable for FlowMRI-Net reconstructions. Namely, such incoherence may be beneficial when exploiting redundancies between velocity encodings for the joint reconstruction [38]. On the other hand, coherence may be beneficial because any undersampling artifacts that remain after reconstruction would be the same for the velocity encodings and hence would disappear when computing phase differences. Future work should investigate a more optimal sampling pattern, possibly incorporating it in the optimization loop for an anatomy-specific sampling [66], [67]. Similarly, the uniform random partitioning, as recommended for multi-mask SSDU [27], and its ratio have not be thoroughly investigated here and may be suboptimal. Although the flow dynamics, physiological motion, SNR, field strength, and vendor of the cerebrovascular 4D flow acquisitions are different compared to the aortic acquisitions, they are both four-point referenced phase-contrast gradient-echo sequences with Cartesian sampling. Future work could investigate FlowMRI-Net's generalizability to different sequence types, such as phase contrast steady-state free precession (PC-SSFP) [68], and non-Cartesian sampling [69]. Finally, a larger cohort including patients should be considered for further validation and proof of generalizability to pathological flow dynamics, such as an intracranial aneurysm, should be investigated.

# Conclusion

FlowMRI-Net is a generalizable deep learning-based reconstruction network that facilitates reconstruction of highly undersampled 4D flow MRI in clinically feasible reconstruction times as exemplarily demonstrated for aortic and cerebrovascular applications. Since FlowMRI-Net is computationally efficient and does not require a reference for training, it can easily be extended to other applications where reference data is not available.

# Data and software availability

The implementation code, data, and pre-trained weights are available at https://gitlab.ethz.ch/ibt-cmr/publications/flowmri_net.

# Acknowledgements

The authors would like to acknowledge financial support through a Personalized Health Related Technology (PHRT) grant of the ETH domain. The provision of the flow WIP package and corresponding support by Drs. Constantin von Deuster, Markus Klarhoefer, and Daniel Giese (Siemens Healthineers) is gratefully acknowledged as well as support by Drs. Patrick Thurner and Zsolt Kulcsar of the Department of Neuroradiology at the University Hospital Zurich.

## Conflict of interest

N/A